%% file: article.tex
\pdfoutput=1
\newif\ifDraft\Drafttrue

\documentclass[letterpaper,10pt,twocolumn]{article}
\usepackage[font={footnotesize,it}]{caption}
\usepackage{usenix}
\usepackage[hyphens]{url}
\usepackage{hyperref}
\usepackage{graphicx}
\usepackage{natbib}
\setcitestyle{numbers,square}

\usepackage{booktabs}

\begin{document}

\date{}
 
\title{This is How You Lose the Transient Execution War}

\author{
	{\rm Allison Randal}\\
	University of Cambridge\\
}

\maketitle

\begin{abstract}
A new class of vulnerabilities related to speculative and out-of-order execution, fault-injection, and microarchitectural side channels rose to attention in 2018. The techniques behind the transient execution vulnerabilities were not new, but the combined application of the techniques was more sophisticated, and the security impact more severe, than previously considered possible. Numerous mitigations have been proposed and implemented for variants of the transient execution vulnerabilities. While Meltdown-type exception-based transient execution vulnerabilities have proven to be tractable, Spectre-type vulnerabilities and other speculation-based transient execution vulnerabilities have been far more resistant to countermeasures. A few proposed mitigations have been widely adopted by hardware vendors and software developers, but combining those commonly deployed mitigations does not produce an effective and comprehensive solution, it only protects against a small subset of the variants. Over the years, newly proposed mitigations have been trending towards more effective and comprehensive approaches with better performance, and yet, older mitigations remain the most popular despite limited security benefits and prohibitive performance penalties. If we continue this way, we can look forward to many generations of hardware debilitated by performance penalties from increasing layers of mitigations as new variants are discovered, and yet still vulnerable to both known and future variants.
\end{abstract}

\input{01_introduction}
\input{02_precursors}
\input{03_spectre}
\input{04_meltdown}

\input{05_beyond}

\input{06_verification}
\input{07_conclusion}

\bibliographystyle{abbrvnat}
\bibliography{article}

\end{document}

%% file: 01_introduction.tex
\section{Introduction}

Early in 2018, two papers by Kocher \textit{et al.} \cite{kocher_spectre_2018} and Lipp \textit{et al.} \cite{lipp_meltdown_2018} and independent work by Google's Project Zero \cite{horn_reading_2018} drew attention to a new class of security vulnerabilities related to both speculative execution and out-of-order execution, collectively described as \textit{transient execution}. The specific vulnerabilities they described---Spectre and Meltdown---use transient execution effects to amplify the severity and ease of exploiting previously known microarchitectural side-channel attacks. Subsequent work has demonstrated that transient execution effects can also be used to amplify the effects of other attacks, such as microarchitectural fault-injection attacks like Rowhammer. The broad class of transient execution vulnerabilities upend traditional notions of secure isolation, and radically expand the potential scope and severity of software-induced hardware vulnerabilities.

The features that the transient execution vulnerabilities exploit are common to modern major hardware architectures, such as x86 and ARM, and had already begun to be replicated in RISC-V implementations before the vulnerabilities were reported, and affect desktop, mobile, embedded, and server hardware. It has been argued that these vulnerabilities are not bugs in the traditional sense, because the transient execution features are functioning as they were designed, however they are flaws in the microarchitecture implementations of both speculative execution and out-of-order pipelines as optimizations to improve instruction-level parallelism. Today, it is possible to mitigate Meltdown-type vulnerabilities in the microarchitecture design with reasonably low performance penalties. Among the major hardware vendors, AMD was never vulnerable to the initial variants of Meltdown \cite{noauthor_speculation_2019, noauthor_software_2023}, and so far it appears that only ARM has made the effort to formally prove that certain generations of their hardware are not vulnerable to Meltdown \cite{madhu_enabling_2022}. Spectre-type vulnerabilities have proven to be more difficult to mitigate, and the products currently shipped by hardware vendors and actively used and deployed around the world offer no more than meager protections---limiting some of the damage caused by some variants, while introducing prohibitive performance penalties---and do not resolve the inherent logic flaws of the microarchitecture implementations, which are the true root cause of the entire class of vulnerabilities.

We can never know what might have happened if the security trade-offs of transient execution had been fully considered at the same time the performance advantages were discovered---whether the transient execution vulnerabilities might have been exposed and resolved earlier, or whether modern computer microarchitectures might have evolved down a slightly different path. If the history of hardware and software security has taught us anything, it is that we have the ability and responsibility to re-consider security trade-offs over time, and make better choices for the future. While it may not be fair to judge past work by lessons we learned later, it will be fair to judge future work on whether it applies those lessons or ignores them.

%% file: 02_precursors.tex
\section{Precursors to Spectre and Meltdown}

While Spectre, Meltdown, and more broadly the entire concept of software-induced transient execution vulnerabilities are relatively new in the field of security research, in essence they are no more than a small step of evolution beyond 70 years of hardware security research on covert channels, side-channel attacks, and fault-injection attacks.

\subsection{Covert channels and side channels}

In 1973, Lampson published ``A note on the confinement problem''  \cite{lampson_note_1973}, an early but influential work on the challenges of preventing information leakage between isolated processes running on the same kernel. In that work he defined a \textit{covert channel} as a hardware resource used to bypass isolation mechanisms by transferring information, where the attack succeeds because the hardware resource was never intended or recognized as a communication channel by the system's designers, so they never bothered to protect it against undesirable information leaks.

Later work uses the term \textit{side channel} in combination with covert channel, but it is important to recognize that although the two terms sometimes appear to be used interchangeably in the literature---and the two kinds of attacks use some of the same hardware resources as channels---covert channels and side channels are not the same thing. In a covert-channel attack, the communication of leaked information is intentional, and the sender and receiver are both malicious (sometimes called ``trojan'' and ``spy''). In a side-channel attack, the communication of leaked information is unintentional, and the sender is a victim, while the receiver is a malicious attacker \cite{ge_survey_2018, szefer_survey_2019, purnal_showtime_2023}.

The hardware resources that Lampson \cite{lampson_note_1973} envisioned being used as covert channels were no more complex than shared memory, a file on the file system, interprocess communication, or request/response metadata, but subsequent work over the decades has explored increasingly exotic channels for leaking information. Conceptually, modern side-channel attacks can trace their roots back to acoustic attacks in the mid-1950s, when recordings of the clicking sounds made by mechanical cryptographic machines captured enough information for attackers to break the cipher used in the encryption \cite{bhunia_hardware_2018, galbally_new_2020}.\footnote{Peter Wright of MI5 \cite[pp.~81-86]{wright_spy_1987} described the attack---later codenamed ENGULF---in Chapter 7 of his autobiography. In 1956, with the help of the London Post Office, he bugged a telephone at the Egyptian Embassy in London installed next to their Hagelin cipher machine, with a hard line to GCHQ so they could listen in each morning as the cipher clerk entered the mechanical encryption settings for the day. Analyzing the recorded sounds with an oscilloscope yielded enough information about how the machine was configured each day that they were able to crack the cipher.} However, there is a world of difference between the 1950s and today in the sophistication of the machines being attacked, the sources of information targeted, the quality and quantity of information gathered from those sources, and the elaborate nature of analysis techniques applied to extract secrets from that information.

\subsection{Physical side-channel attacks}

The first rounds of research into side channels focused on physical side-channel attacks, exploiting indirect physical information to extract secrets. Because physical side-channel attacks require physical access or proximity to the machine, they are more difficult to perform, and have historically been regarded as less risky and only worth mitigating on security-critical components such as cryptographic hardware. The most common kinds of physical information gathered in these attacks, which still remain relevant today, are:

\begin{itemize}
	
	\item Timing Analysis: measures execution time of operations (such as encryption/decryption) for different inputs, and infers secret information from variations in timing. This technique is often combined with other physical side-channel attacks. In the mid-1990s, Kocher \cite{kocher_timing_1996} advanced this technique---and the entire research field of physical side-channel attacks---to a point of being able to extract entire secret keys from a decryption process.
	
	\item Power Analysis: measures power usage related to operations (such as encryption/decryption) for different inputs/outputs, and
	infers secret information from variations in power
	consumption. In the late 1990s, Kocher \textit{et al.} \cite{kocher_differential_1999} made similar advances in physical side-channel attack techniques making use of power analysis.
	
	\item Electromagnetic Analysis: measures electromagnetic waves produced by current flow over the device, and infers secret information from variations in electromagnetic signals. In the early 2000s, Quisquater and Samyde \cite{quisquater_electromagnetic_2001} built on Kocher's earlier work on timing analysis and power analysis to extract secret keys from smart cards using only electromagnetic analysis.
	
	\item Fault Analysis: physically tampers with voltage levels, clock signal, or other hardware components to trigger a fault in the device (e.g. disturb a few memory or register bits), and infers secret information based on variations in the output of faulty operations. This is actually a combination of two techniques, it starts with a physical fault-injection attack (violating integrity), then uses the successful results of the fault-injection attack as a source of information for a physical side-channel attack (violating confidentiality). In the mid-1990s, Anderson and Kuhn \cite{anderson_tamper_1996} made a first brief mention of clock and power glitching techniques in the context of smart card attacks, which Skorobogatov and Anderson \cite{skorobogatov_optical_2003} later explicitly connected with Kocher's work on physical side-channel attacks.

\end{itemize}

\subsection{Microarchitectural side-channel attacks}\label{sec:uarch-side-channels}

More recent rounds of research into side-channel attacks have expanded the range of information sources considered. In contrast to physical side-channel attacks, microarchitectural side-channel attacks exploit indirect microarchitectural sources of information to extract secrets, do not require physical access to the machine, and may even be software-induced, so they are easier to perform and of greater concern for general-purpose hardware. The analysis techniques and objectives of microarchitectural side-channel attacks are similar to earlier work on physical side-channel attacks, though the sources of information used are more varied and also inspired by that earlier work.

As a common example of microarchitectural side-channel attack techniques, cache-timing analysis measures the time required to load a data value from cache, and infers secret information from variations in timing. An attacker establishes a pre-defined cache state, allows the victim to perform an operation, then observes cache state changes. Although Kocher \cite{kocher_timing_1996} briefly mentioned the influence cache-timing effects have on physical timing analysis in the mid-1990s, the idea of microarchitecture cache-timing side-channel attacks was not fully developed until the mid-2000s by Bernstein \cite{bernstein_cache-timing_2004} and Percival~\cite{percival_cache_2005}, who extracted entire secret keys using only cache-timing information. Cache-timing side-channel attacks have been a prolific area of security research for nearly two decades, with variants differentiated by characteristics like the specific cache targeted (for an L1 attack to succeed, the attacker and victim have to share a core, while an LLC attack can succeed across cores), or the specific attacker actions to prepare or observe the cache, such as Prime+Probe \cite{osvik_cache_2006, liu_last-level_2015}, Evict+Time \cite{osvik_cache_2006}, Flush+Reload \cite{yarom_flush+reload:_2014}, Flush+Flush \cite{gruss_flushflush_2016}, Stream+Reload \cite{weber_osiris_2021}, or Write+Write \cite{thoma_write_2022}.

Caches are not the only targets for microarchitectural side-channel attacks, many other microarchitectural sources of information have been successfully exploited to extract secrets, such as:

\begin{itemize}
	
	\item Translation Lookaside Buffer (TLB): Wang \textit{et al.} \cite{wang_leaky_2017}, TLBleed \cite{gras_translation_2018} successfully bypasses cache isolation
	\item Page tables: Van Bulck \cite{van_bulck_telling_2017}, Wang \textit{et al.} \cite{wang_leaky_2017}
	\item DRAM: Pessl \cite{pessl_drama_2016}, Wang \textit{et al.} \cite{wang_leaky_2017}
	\item Prefetchers: Szefer \cite{szefer_survey_2019}, Shin \textit{et al.} \cite{shin_unveiling_2018}, Vicarte \textit{et al.} \cite{sanchez_vicarte_opening_2021}, AfterImage \cite{chen_afterimage_2023}
	\item Branch Target Buffer (BTB): Branch Prediction Analysis (BPA) \cite{aciicmez_predicting_2006} and Simple Branch Prediction Analysis (SBPA) \cite{aciicmez_power_2007}, Evtyushkin \textit{et al.} \cite{evtyushkin_jump_2016}, Lee \textit{et al.} \cite{lee_inferring_2017}, Yu \textit{et al.} \cite{yu_all_2023}
	\item Conditional branch predictor, Pattern History Table (PHT): BranchScope \cite{evtyushkin_branchscope_2018}, Bluethunder \cite{huo_bluethunder_2020}
	\item Return Stack Buffer (RSB): Hyper-Channel \cite{bulygin_cpu_2008}
	\item FPU timing: Andrysco \textit{et al.} \cite{andrysco_subnormal_2015}
	\item SMT port contention: Wang and Lee \cite{wang_covert_2006}, Aciicmez and Seifert \cite{aciicmez_cheap_2007}, Aldaya \textit{et al.} \cite{aldaya_port_2019}
	\item GPU timing: Xu \textit{et al.} \cite{xu_gpuguard_2019}
	\item CPU frequency: CLKscrew \cite{tang_clkscrew_2017}
	\item Power analysis: Hertzbleed \cite{wang_hertzbleed_2022}, Platypus \cite{lipp_platypus_2021}, Barenghi and Pelosi \cite{barenghi_side-channel_2018}, Collide+Power \cite{kogler_collidepower_2023}
	\item Memory controller scheduler: Semal \textit{et al.} \cite{semal_leaky_2020}
	\item Cache way predictor: Take A Way \cite{lipp_take_2020}
	\item Instruction cache: Aciiçmez \cite{aciicmez_yet_2007}, Aciiçmez \textit{et al.} \cite{aciicmez_new_2010}
	\item Micro-op cache: Ren \textit{et al.} \cite{ren_i_2021}
	\item Performance counters: PMU-Leaker \cite{qiu_pmu-leaker_2023}

\end{itemize}

Spectre and Meltdown build on this history of research into side-channel attacks. They make use of microarchitectural side-channel attack techniques, but are often falsely categorized simply as timing analysis techniques, specifically as cache-timing side-channel attacks \cite{canella_evolution_2020}. It is more accurate to recognize that Spectre and Meltdown are both primarily fault analysis techniques, because they both begin with a fault-injection attack (violating integrity)---Meltdown by triggering an exception, and Spectre by inserting false entries into branch prediction and other prediction-related microarchitectural state---and then go on to use the successful results of the microarchitectural fault-injection attack as a source of information for a microarchitectural side-channel attack (violating confidentiality).

\subsection{Transient execution}

The concepts of transient execution, transient instructions, and transient microarchitectural state are modern terminology to describe some curious side-effects of the way general-purpose high-performance processors have been designed since the 1960s. The main features of concern for transient execution are speculative and out-of-order execution, but transient execution effects are compounded by the interactions between several features, including multilevel memory caches, simultaneous multithreading, multiple instruction issue, and prefetching.

In the late 1960s, Tomasulo \cite{tomasulo_efficient_1967} discussed an approach to dynamically scheduling the execution of instructions across multiple execution units, as implemented for floating-point operations in the IBM 360/91. The key insight of the approach was that instructions could be reordered from the original program sequence, as long as dependencies between instructions were preserved. One usability problem with this early implementation of out-of-order execution was that it delivered interrupts chaotically out of order too, because it had no concept of a separate in-order commit stage, and simply committed instructions as soon as they finished executing \cite{patterson_computer_2017}. So, later implementations of out-of-order execution delayed interrupts and exceptions until an in-order commit stage, so they would be delivered in program order.

A collection of papers in the early 1970s, including Tjaden and Flynn \cite{tjaden_detection_1970}, Flynn \cite{flynn_computer_1972}, Flynn and Podvin \cite{flynn_shared_1972}, and Riseman and Foster \cite{riseman_inhibition_1972}, explored the logical limits of instruction-level parallelism for the hardware of the time, identifying branches and memory loads as significant obstacles. Within a decade, the tone of publications shifted from assessing these obstacles as insurmountable, to assessing them as straightforwardly solved by combining several techniques that remain in common use today, particularly the speculative techniques of branch prediction and memory load prediction. Lee and Smith \cite{lee_branch_1984} and McFarling and Hennessy \cite{mcfarling_reducing_1986} captured historical perspectives on branch prediction from the point of view of the mid-1980s. Both surveyed the state of the art in branch prediction techniques at the time---such as dynamic prediction and branch target buffers---and critically reviewed previous techniques to speed up conditional branches without speculation---such as delayed branches, look-ahead resolution, branch target prefetching, and multiple instruction streams.

One noteworthy characteristic shared by these early papers---and by much of the substantial work on speculative and out-of-order pipeline techniques in the decades that followed---was a focus on metrics of performance with little or no consideration given to metrics of security. In all fairness to the hardware designers of the time, the groundbreaking work on speculation and out-of-order execution was completed decades before microarchitectural side-channel attacks were considered as a possibility. So, their oversight was not a matter of willfully ignoring known threats, it was a naive complacency and unsophisticated design methodology that embraced new features without adequate consideration of the system-wide implications. Modern hardware designers have no such excuse. Some of the earliest work on microarchitectural side-channel attacks in the mid-2000s by Percival~\cite{percival_cache_2005} explored the risks inherent in combining speculative execution with simultaneous multithreading, dynamic pipeline scheduling, multilevel memory caches, and hardware prefetching---identifying the essential constituents of the transient execution vulnerabilities over a decade before the full extent of their security impact was revealed. Fogh~\cite{fogh_negative_2017} also identified the potential risk that speculative and out-of-order execution could be used to amplify microarchitectural side-channel attacks in 2017, but did not formulate a successful attack.

%% file: 03_spectre.tex
\section{Spectre}

Spectre is a hardware security vulnerability first discovered in 2017, but not reported publicly until January 2018 by Kocher \textit{et al.} \cite{kocher_spectre_2018}. Together with Meltdown, Spectre is the first of a new class of vulnerabilities---known as transient execution vulnerabilities---that exploit weaknesses in certain low-level microarchitectural effects of out-of-order and speculative pipelines. While any out-of-order pipeline could be vulnerable to Meltdown, only speculative pipelines can be vulnerable to Spectre. Spectre is a fault analysis side-channel attack---it combines both fault-injection techniques to manipulate the victim into a vulnerable state and side-channel techniques to convey the exposed secrets to the attacker. The combination of the two techniques is what makes this class of vulnerabilities so powerful. The fault-injection phase of a Spectre-type attack mistrains a speculative predictor so it starts making false predictions. The victim blindly accepts the false predictions and proceeds to execute with either wrong values or wrong instructions, leaving a trail of microarchitectural state changes as it executes. In theory, those microarchitectural state changes are architecturally invisible if the speculated prediction proves to be false---the architectural changes are all cleaned away and the pipeline is flushed leaving no visible effects\footnote{Some architectures are sloppier than others about cleaning up the side-effects speculation.}---so they are ``transient'' in the sense that they only exist briefly before they disappear \cite{canella_systematic_2019, canella_evolution_2020}. But, during transient execution, the microarchitectural state changes that the victim made as a result of false predictions are microarchitecturally visible, so the attacker can access them through side channels. All the side channels used as the transmission phase of Spectre-type attacks can be used as stand-alone microarchitectural side-channel attacks, and as we discussed above in Section \ref{sec:uarch-side-channels}, some of those side channels have been known for decades. The uniquely interesting thing about Spectre is the initial fault-injection preparation phase, which tricks the victim into exposing its own secrets---the attacker manipulates the victim into executing instructions or values it never would have done non-speculatively, so the victim creates shared microarchitectural state it never would have created non-speculatively, specifically so the attacker can access that shared microarchitectural state through microarchitectural side channels.

\subsection{Characterizing the variants}

The first few variants of Spectre published early in 2018 were novel, but also relatively simple. After 6 years and hundreds of published papers, the landscape today is a combinatorial explosion of variants and mitigations. Understanding the first few variants published is not enough to make sense of the entire class of Spectre-type vulnerabilities, but far too many hardware researchers and engineers make the mistake of stopping there.

The discouraging truth of Spectre is that potentially any speculative predictor could be used for the fault-injection preparation phase, and potentially any microarchitectural state could be used for the side-channel transmission phase. To compound the complexity, in the access phase any instructions executed transiently by the victim as a result of the fault-injection misprediction could take any action to leave transient traces in shared microarchitectural state, serving as a \textit{gadget} that exposes secrets so they become vulnerable to side-channel transmission. All of those variations in the preparation, access, or transmission phases are still called ``Spectre'', because they all satisfy the fundamental definition of the technique---as Kocher \textit{et al.} \cite{kocher_spectre_2018} described it in the very first paper, ``Spectre attacks involve inducing a victim to speculatively perform operations that would not occur during correct program execution and which leak the victim's confidential information via a side channel to the adversary.''

The primary way of categorizing Spectre-type variants, shown in Table \ref{table:spectre-variants-by-predictor}, is by the fault-injection attack vector used to trigger speculative execution in the preparation phase. While the initial Spectre variants reported in 2018 used a branch, return, or memory dependence predictor in the preparation phase, subsequent work on Spectre and other transient execution vulnerabilities has explored a more diverse collection of ways to trigger speculative execution in the pipelines of modern processors, as shown in Table \ref{table:spectre-variants-by-predictor} and further discussed in Section \ref{sec:beyond-spectre-meltdown}. The choice of predictor in the preparation phase has a significant impact on later phases of a Spectre attack. For example, there is a fundamental difference between Spectre variants with an attack vector of conditional branch prediction and Spectre variants with an attack vector of direct or indirect branch prediction, or return prediction. In some ways conditional branch predictors are less powerful attack vectors, because their control flow destinations are limited to two alternatives---either redirecting control flow to one specific label or continuing to the next instruction---rather than being able to redirect control flow to an arbitrary mispredicted address. But, conditional branch predictors also make predictions about the value evaluated by the condition, and that predicted value can be used in later phases of the attack. Some Spectre variants depend on a wrong value prediction, while other variants work equally well with any control flow predictor.

\begin{table*}[h!]
	\small
	\caption{Spectre variants by preparation phase fault-injection attack vector} \label{table:spectre-variants-by-predictor}
	\begin{tabular}{p{4cm} p{5cm} p{6cm}}
		\toprule
		\textbf{Predictor} & \textbf{Mechanisms} & \textbf{Examples} \\
		\midrule

		Pattern History Table (PHT) or Conditional Branch Predictor (CBP) & PHT/CBP poisoning: mistrains conditional branch prediction, to redirect control flow to the attacker's chosen branch path, so the victim transiently executes either wrong instructions or with wrong values. & Spectre-PHT (Spectre variant 1, ``Input Validation Bypass'') \cite{kocher_spectre_2018}, Kiriansky and Waldspurger (Spectre variants 1.1 and 1.2) \cite{kiriansky_speculative_2018}, NetSpectre \cite{schwarz_netspectre:_2018}, SGXSpectre \cite{okeeffe_spectre_2018}, SiSCloak \cite{buiras_validation_2021}, HammerScope \cite{cohen_hammerscope_2022}, SpecHammer \cite{tobah_spechammer_2022}, Schwarzl \textit{et al.} \cite{schwarzl_robust_2022}\\
		\midrule

		Branch Target Buffer (BTB) & BTB poisoning: mistrains direct or indirect branch prediction, to redirect control flow to the attacker's chosen branch destination, so the victim transiently executes wrong instructions. & Spectre-BTB (Spectre variant 2, ``Branch Target Injection'') \cite{kocher_spectre_2018,noauthor_retpoline:_2018,vougioukas_brb_2019}, SgxPectre \cite{chen_sgxpectre_2019}, Spectre-BTB-SA-IP \cite{canella_systematic_2019}, SMoTherSpectre \cite{bhattacharyya_smotherspectre_2019},  Mambretti \textit{et al.} \cite{mambretti_two_2019}, Straight-Line Speculation (BTB variants) \cite{noauthor_straight-line_2020}, Retbleed \cite{wikner_retbleed_2022}\\
		\midrule

		Branch History Buffer (BHB) & BHB poisoning: mistrains indirect branch prediction, to redirect control flow to the attacker's chosen branch destination, so the victim transiently executes wrong instructions. & Spectre-BHB (``Branch History Injection'') \cite{barberis_branch_2022}\\
		\midrule

		Return Stack Buffer (RSB) or Return Address Stack (RAS) & RSB poisoning: mistrains the RSB by executing call instructions to add invalid entries to the RSB, or explicitly overwrites return addresses, to redirect return control flow to the attacker's chosen destination, so the victim transiently executes wrong instructions. & Spectre-RSB (Spectre variant 5, ``Return Address Injection'') \cite{maisuradze_ret2spec_2018, koruyeh_spectre_2018}, SgxPectre (RSB falls back on BTB) \cite{chen_sgxpectre_2019}, Straight-Line Speculation (RSB variants) \cite{noauthor_straight-line_2020}, Spring \cite{wikner_spring_2022}, Inception \cite{trujillo_inception_2023} \\
		\midrule

		Memory dependence predictor & STL poisoning: mistrains store-to-load predictor, so the victim transiently loads stale values that should have been overwritten by intervening stores, and transiently executes with wrong values. If the stale value is a code pointer, it can redirect control flow to a gadget, so the victim transiently executes the wrong instructions. & Spectre-STL (Spectre variant 4, ``Speculative Store Bypass'') \cite{horn_speculative_2018}\\
		\midrule
		
		String Comparison Overrun (SCO) & Does not require mistraining or a leakage gadget, because a single instruction contains both the speculation trigger and the leaking memory access & Oleksenko \textit{et al.} \cite{oleksenko_hide_2023} \\
		\midrule
		
		Zero Dividend Injection (ZDI) & Speculation induced by division instructions & Oleksenko \textit{et al.} \cite{oleksenko_hide_2023} \\
		\bottomrule
		
	\end{tabular}
\end{table*}

As discussed in Section \ref{sec:uarch-side-channels}, many different microarchitectural states have been exploited in microarchitectural side-channel attacks. So, it should come as no surprise that the side-channel attack vectors used in the transmission phase of Spectre-type vulnerabilities have been equally diverse, some of the highlights are listed in Table \ref{table:spectre-variants-by-channel}. Not every microarchitectural side channel listed in Section \ref{sec:uarch-side-channels} has a corresponding paper demonstrating that it can be exploited in a Spectre-type variant, and new microarchitectural side channels are still being discovered, so the list in Table \ref{table:spectre-variants-by-channel} continues to grow. Over time, while new research publications continue to explore individual side channels to discover new Spectre-type variants, there is also a growing body of research into developing tools to find side channels that can be exploited by Spectre-type variants and other transient execution vulnerabilities, as discussed in Section \ref{sec:verification}.

\begin{table*}[h!]
	\small
	\caption{Spectre variants by transmission phase side-channel attack vector} \label{table:spectre-variants-by-channel}
	\begin{tabular}{p{4cm} p{5cm} p{5cm}}
		\toprule
		\textbf{Channel} & \textbf{Mechanisms} & \textbf{Examples} \\
		\midrule
		
		L1 data cache & Leaks information using a cache-timing side channel on the L1D cache & Take A Way \cite{lipp_take_2020}\footnote{While Lipp \textit{et al.} \cite{lipp_take_2020} demonstrated that AMD's cache way predictor can be used as part of a Spectre variant, it is only used as a side-channel attack vector in the transmission phase, not as a fault-injection attack vector in the preparation phase.},  PMU-Leaker \cite{qiu_pmu-leaker_2023}, most attack variants that succeed with L3 as a side channel also work on L1D\\
		\midrule
		
		L1 instruction cache & Leaks information using a cache-timing side channel on the L1I cache & Mambretti \textit{et al.} \cite{mambretti_two_2019}\\
		\midrule

		L2 cache & Leaks information using a cache-timing side channel on the L2 cache & Most attack variants that succeed with L3 as a side channel also work on L2\\
		\midrule

		L3/Last-level cache & Leaks information using a cache-timing side channel on the L3 cache or LLC, for example, Flush+Reload \cite{yarom_flush+reload:_2014} or Prime+Probe \cite{liu_last-level_2015} & SgxPectre \cite{chen_sgxpectre_2019}\\
		\midrule
		
		Translation Lookaside Buffer (TLB) & Leaks information using a TLB-based side channel & Yan \textit{et al.} \cite{yan_invisispec_2018}, Khasawneh \textit{et al.} \cite{khasawneh_safespec_2019}, Kiriansky \textit{et al.} \cite{kiriansky_dawg_2018}, Loughlin \textit{et al.} \cite{loughlin_dolma_2021}, Schwarz \textit{et al.} \cite{schwarz_store--leak_2021}, Seddigh \textit{et al.} \cite{seddigh_breaking_2022}, PACMAN \cite{ravichandran_pacman_2022} \\
		\midrule
		
		Vector instructions & Leaks information using a side channel based on differences in AVX2 instruction timing  & NetSpectre \cite{schwarz_netspectre:_2018}, Weber \textit{et al.} \cite{weber_osiris_2021}\\
		\midrule
		
		SMT and single-threaded port contention & Leaks information using a side channel based on execution timing differences between instructions on different execution ports & SMoTherSpectre \cite{bhattacharyya_smotherspectre_2019}, Fustos \textit{et al.} \cite{fustos_spectrerewind_2020}, Spectre-STC \cite{fadiheh_exhaustive_2023}\\
		\midrule
		
		Branch Target Buffer (BTB) & Leaks information using a side channel based on timing differences between correct and false BTB predictions\footnote{As noted in Table \ref{table:spectre-variants-by-predictor}, the BTB can also be used as a fault-injection attack vector in the preparation phase of Spectre.} & Weisse \textit{et al.} \cite{weisse_nda:_2019}, Mambretti \textit{et al.} \cite{mambretti_two_2019} \\
		\midrule
		
		Micro-op cache & Leaks information using a micro-op cache-timing side channel & Ren \textit{et al.} \cite{ren_i_2021} \\
		\midrule
		
		Instruction timing & Leaks information using a side channel based on variable-time arithmetic instructions & Zhang \textit{et al.} \cite{zhang_ultimate_2023}, Rajapksha \textit{et al.} \cite{rajapaksha_sigfuzz_2023} \\
		\midrule
		
		Store and load buffers & Leaks information using a side channel based on execution timing analysis of load-store buffers & Timed Speculative Attacks (TSA) \cite{chakraborty_timed_2022} \\
		\midrule
		
		Rowhammer & Leaks information using a side channel based on measuring the power consumed by transient memory accesses & HammerScope \cite{cohen_hammerscope_2022} \\
		\midrule
		
		Performance Monitor Unit (PMU) & Leaks information using a side channel based on performance counters & PMU-Spill \cite{qiu_pmu-spill_2023} \\
		\bottomrule

	\end{tabular}
\end{table*}

\subsection{Characterizing the countermeasures}

\begin{figure}[h]
	\centering
	\includegraphics[scale=0.45]{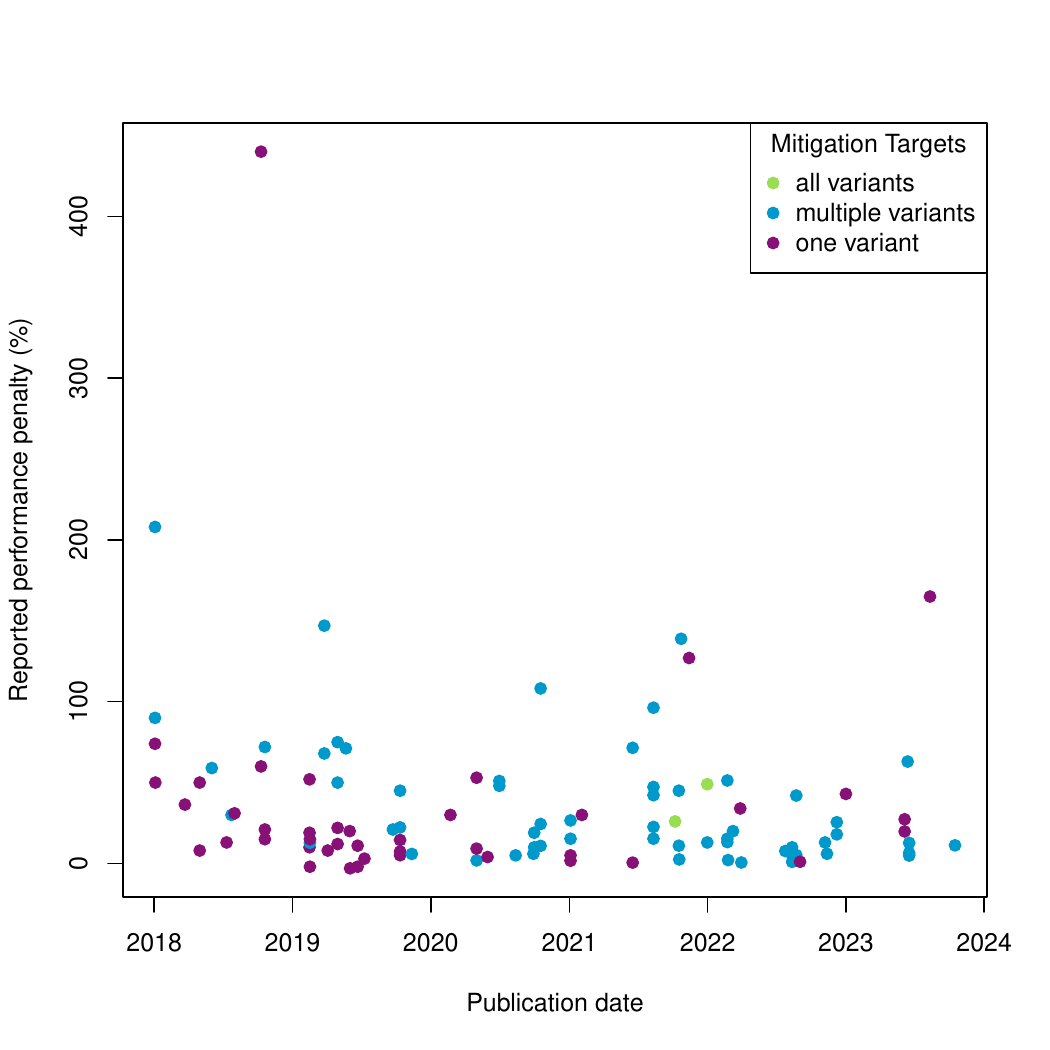}
	\caption{Performance penalty trends for Spectre countermeasures (2018-2023)}
	\label{fig:plot_spectre_mitigations}
\end{figure}

Many countermeasures for Spectre-type vulnerabilities have been proposed, but overall the results have been disappointing \cite{canella_evolution_2020, fiolhais_transient-execution_2023}. As Figure \ref{fig:plot_spectre_mitigations} illustrates,\footnote{The data sources for Figures \ref{fig:plot_spectre_mitigations} and \ref{fig:plot_spectre_hwsw} are \cite{ainsworth_ghostminion_2021,ainsworth_muontrap_2020,amit_jumpswitches_2019,bahmani_cure_2021,barber_specshield_2019,behrens_performance_2022,bourgeat_mi6_2019,canella_systematic_2019,carruth_introduce_2018,carruth_speculative_2018,choudhari_specdefender_2022,choudhary_speculative_2021,dessouky_hybcache_2020,dong_spectres_2018,escouteloup_under_2022,fustos_spectreguard_2019,ge_more_2022,gonzalez_replicating_2019,green_safebet_2023,hertogh_quarantine_2023,jin_specterminator_2022,khasawneh_safespec_2019,kim_revice_2020,kiriansky_dawg_2018,kocher_spectre_2018,koruyeh_speccfi_2020,kuzuno_mitigating_2022,kvalsvik_doppelganger_2023,larabel_bisected_2018,le_cross-process_2023,lee_cacherewinder_2022,li_conditional_2019,li_conditional_2022,li_fase_2022,loughlin_dolma_2021,mcilroy_spectre_2019,narayan_swivel_2021,ojogbo_secure_2020,oleksenko_you_2018,patrignani_exorcising_2021,randal_ghosting_2021,reis_mitigating_2018,sabbagh_secure_2021,saileshwar_cleanupspec_2019,sakalis_efficient_2019,sakalis_ghost_2019,schwarz_context_2019,schwarzl_specfuscator_2021,shen_restricting_2019,taram_context-sensitive_2019,taram_secsmt_2022,thoma_basicblocker_2021,tkachenko_20-30_2018,tran_clearing_2020,vassena_automatically_2021,vougioukas_brb_2019,wang_oo7_2019,wang_svp_2023,weisse_nda:_2019,wistoff_microarchitectural_2021,wistoff_systematic_2022,wu_rcp_2022,wu_reversispec_2020,yan_invisispec_2018,yu_speculative_2019,yu_speculative_2020,zhang_ultimate_2023,zhao_pinned_2022,zhao_speculation_2020}. Some performance results are self-reported, while others are reported by subsequent papers evaluating earlier papers.} the performance penalties of proposed mitigations have been improving over time, and the proposals are trending toward mitigating more than one variant by considering root causes. Unfortunately, it is relatively common to see papers---such as Behrens \textit{et al.} \cite{behrens_performance_2022} or Guan \textit{et al.} \cite{guan_building_2019}---which claim to evaluate the overall performance of mitigating Spectre, but actually only evaluate a small subset of mitigations that are inadequate to mitigate all variants. So far, the only approach that eliminates all variants of Spectre is to eliminate speculation entirely, and while the approach is often dismissed for performance reasons without any actual performance measurements \cite{kocher_spectre_2018, lipp_meltdown_2018, schwarz_netspectre:_2018, yan_invisispec_2018, canella_systematic_2019, gonzalez_replicating_2019, sakalis_efficient_2019, he_new_2021, weisse_nda:_2019, rokicki_ghostbusters_2020}, the few papers that do measure the performance of eliminating speculation \cite{thoma_basicblocker_2021, randal_ghosting_2021} reveal performance penalties comparable to other mitigations for Spectre.\footnote{The two green ``all variants'' data points in Figure \ref{fig:plot_spectre_mitigations} are both non-speculative.}

\subsubsection{Software-only mitigation approaches}

Some of the earliest mitigations proposed for Spectre were software workarounds for the vulnerabilities. These mitigations were inspired by earlier work on mitigating side-channel attacks for cryptographic software, where it was understood that the mitigations only needed to be applied to small but critical sections of code \cite{coppens_practical_2009, ge_survey_2018, szefer_survey_2019, lou_survey_2021, moghimi_memjam_2019, canella_evolution_2020, cauligi_constant-time_2020, bognar_microprofiler_2023}. Software-only mitigations have the advantage that they require no changes to the hardware, however, they have prohibitive performance penalties, and have proven to be inconsistently effective.

The very first paper on Spectre by Kocher \textit{et al.} \cite{kocher_spectre_2018} suggested the insertion of speculation barrier instructions---for example, \texttt{lfence} on x86 or \texttt{sb} on ARM (added in v8.0)---which temporarily block speculative execution for instructions after the barrier, until speculation has resolved for all instructions before the barrier. The major vendors quickly adopted this approach and still actively recommended it today \cite{noauthor_software_2023, noauthor_speculative_2021}. Oleksenko \textit{et al.} \cite{oleksenko_you_2018} demonstrated performance penalties as high as 440\% for comprehensive use of \texttt{lfence}, which is worse than simply eliminating speculation \cite{canella_evolution_2020}. The focus of much subsequent work on speculation barriers has been on limiting their use to improve performance \cite{wang_oo7_2019, taram_context-sensitive_2019, vassena_automatically_2021, johannesmeyer_kasper_2022}. However, anything less than comprehensive use of speculation barriers means there is no guarantee that Spectre is fully mitigated \cite{silberstein_speculating_2018, taram_context-sensitive_2019, vassena_automatically_2021}. Manual placement of speculation barriers is prone to developer mistakes, automatic placement often misses vulnerable code patterns, and even when the speculation barriers are correctly placed, race conditions in the specific microarchitecture implementation may allow secrets to be leaked past the barrier anyway \cite{milburn_you_2022, ojogbo_secure_2020}. As with many Spectre mitigations, speculation barriers are often targeted only at the most well-known variants, and fail to provide protection beyond that narrow scope. For example, \texttt{lfence} is not effective against Spectre variants that use alternative side-channels as the transmission phase of the attack, such as side-channels based on AVX functional units, the TLB, the instruction cache \cite{schwarz_netspectre:_2018}, or micro-op cache \cite{ren_i_2021}, or against Spectre variants that use a speculative write to modify the gadget code \cite{kiriansky_speculative_2018}. Intel added a new Indirect Branch Predictor Barrier (IBPB) \cite{noauthor_speculative_2018} instruction in 2018 to manually flush indirect branch predictor state so branch predictions after the barrier are not trained by branches before the barrier at a performance penalty of 24\% to 53\% \cite{koruyeh_speccfi_2020}, but Wikner and Ravazi \cite{wikner_retbleed_2022} demonstrated that this mitigation was incomplete.

Another early software-only mitigation for Spectre-BTB was retpoline \cite{turner_retpoline_2018, noauthor_retpoline:_2018}, which replaces an indirect branch instruction with a return sequence in the instruction stream. McIlroy \textit{et al.} \cite{mcilroy_spectre_2019} reported a performance penalty of 152\% for comprehensive use of retpoline, and subsequent work has focused on limiting the use of retpoline \cite{koruyeh_speccfi_2020, joly_evaluation_2020}. Initially, retpoline was constructed on the assumption that the Return Stack Buffer could not be mistrained by attackers, but the Spectre-RSB variant \cite{maisuradze_ret2spec_2018, koruyeh_spectre_2018} later proved that assumption to be false and bypassed retpoline as a mitigation for Spectre-BTB. Maisuradze and Russow \cite{maisuradze_ret2spec_2018} suggested an alternative form of retpoline as a mitigation for the Spectre-RSB variant. The Retbleed \cite{wikner_retbleed_2022} variant of Spectre demonstrated that retpoline is not an effective mitigation on architectures such as Intel and AMD that fallback to the Branch Target Buffer (BTB) to predict returns.

Speculative Load Hardening (SLH) is another software mitigation technique, which only mitigates the Spectre-PHT variant, proposed by Carruth \cite{carruth_speculative_2018} in 2018, adopted by both LLVM and GCC, with a reported performance penalty of 36\% \cite{canella_systematic_2019}. In 2021, Patrignani and Guarnieri \cite{patrignani_exorcising_2021} demonstrated that the original implementation of Speculative Load Hardening still allowed some data leaks, and proposed a stronger form of the mitigation, with a reported performance penalty of 127\% \cite{zhang_ultimate_2023}. In 2023, Zhang \textit{et al.} \cite{zhang_ultimate_2023} demonstrated that the original SLH mitigation is not effective against alternative side-channels in the transmission phase based on variable-time arithmetic instructions, and proposed an improved ``ultimate'' SLH mitigation, with a reported performance penalty of 165\%.

Swivel \cite{narayan_swivel_2021} applied compiler transformations to sandboxed WASM code to limit some of the effects of Spectre vulnerabilities, however the approach relies on techniques like fences, ASLR, BTB flushing, and Intel's MPK which have been demonstrated not to be effective \cite{schwarz_netspectre:_2018, kiriansky_speculative_2018, canella_systematic_2019, canella_evolution_2020, ren_i_2021, wikner_retbleed_2022, seddigh_breaking_2022}. Several authors pointed out that Swivel and other compiler-based mitigations such as Jenkins \textit{et al.} \cite{jenkins_ghostbusting_2020} and Venkman \cite{shen_restricting_2019}, have never been verified to work \cite{choudhari_specdefender_2022, cauligi_sok_2022, zhang_ultimate_2023}.

McIlroy \textit{et al.} \cite{mcilroy_spectre_2019} noted that in their analysis, it was not possible to address the Spectre-STL variant using software-only mitigations.

\begin{figure}[h]
	\centering
	\includegraphics[scale=0.45]{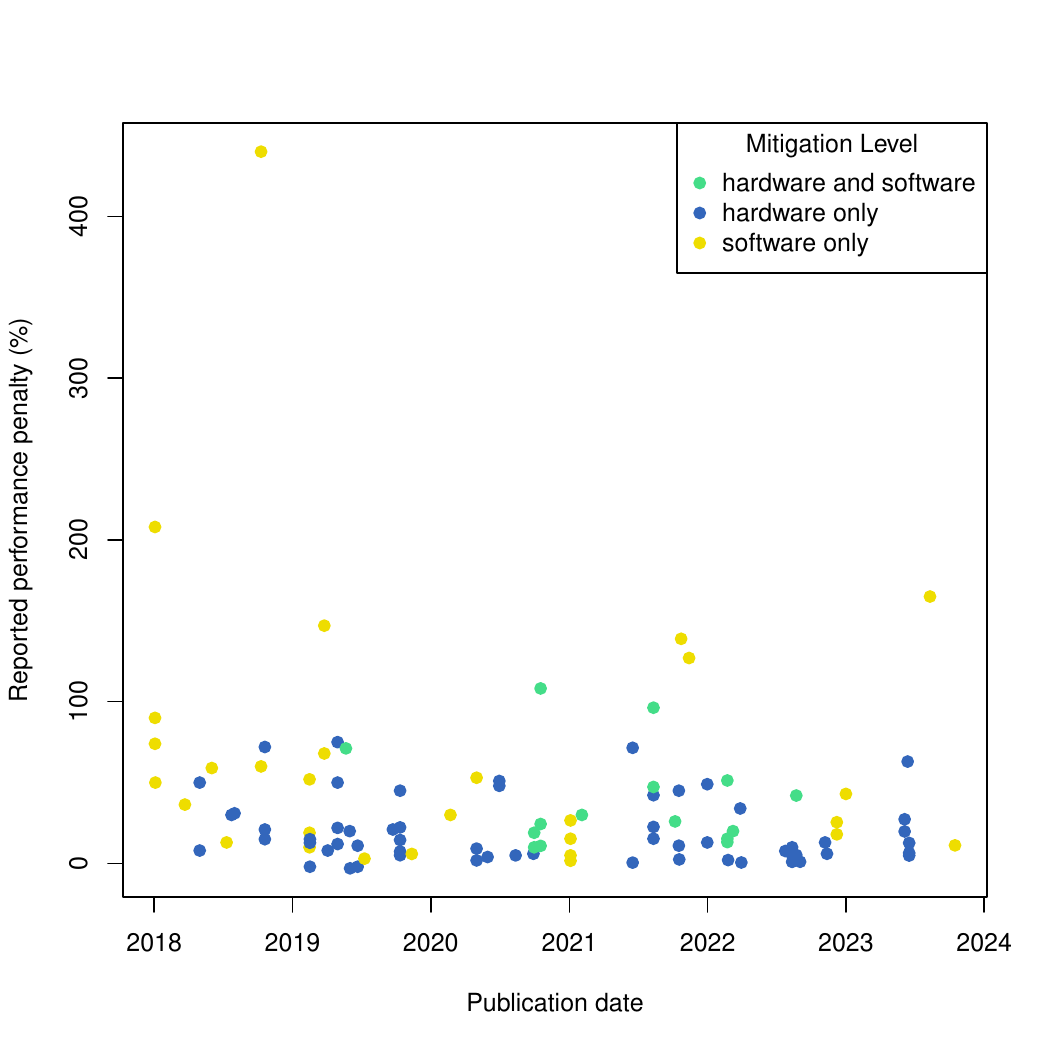}
	\caption{Performance penalty trends by implementation level for Spectre countermeasures (2018-2023)}
	\label{fig:plot_spectre_hwsw}
\end{figure}

While the initial mitigations proposed for Spectre were mostly implemented as software patches, over the years the trend has shifted toward mitigations implemented entirely in hardware or with an element of hardware acceleration, as shown in Figure \ref{fig:plot_spectre_hwsw}. One factor in the decline of software-only mitigations is that hardware mitigations have tended to perform better than software mitigations. Another factor is that historically, hardware architectures were rarely designed with the intention of giving software control over speculation features,\footnote{The Intel i860 \cite{kohn_introducing_1989} was one noteworthy exception.} so the range of options for mitigating Spectre entirely in software have been limited. The software mitigations proposed in recent years have often been refinements of software mitigations from previous years, such as successive attempts to improve the security of Speculative Load Hardening (SLH) \cite{patrignani_exorcising_2021, zhang_ultimate_2023} or to improve the performance of fences \cite{wang_oo7_2019, zhao_speculation_2020, vassena_automatically_2021, zhao_pinned_2022, wistoff_systematic_2022}.

\subsubsection{Mitigation approaches that only consider cache-based side-channels}

It is unfortunately common for papers about Spectre to focus on variants of the vulnerability that use cache-based side-channels, and then propose cache-based mitigations as if they could be solutions for Spectre. Some early papers went so far as to classify Spectre simply as a cache-timing side-channel attack without any mention of the transient execution effects involved \cite{chattopadhyay_symbolic_2018, qureshi_ceaser_2018}. Such an oversimplification of Spectre-type vulnerabilities indicates a lack of understanding of past work and the available literature on Spectre. Even the very first paper on Spectre by Kocher \textit{et al.} \cite{kocher_spectre_2018} explicitly discussed the fact that many different microarchitectural side-channels could be used for the transmission phase of Spectre, though the specific examples they chose to implement for the paper used cache-timing side-channels. While it is worthwhile to review these mitigation proposals as part of a comprehensive survey on Spectre, it is also important to recognize that exclusively cache-based mitigations can never be anything more than partial solutions \cite{weisse_nda:_2019, chakraborty_timed_2022}.

One group of papers in this category are really no more than general mitigations for cache-timing side-channel attacks. Although they mention Spectre (and sometimes also Meltdown) vulnerabilities as prominent examples, they do not make specific claims that their approach is a viable one for transient execution vulnerabilities. For example, CEASER \cite{qureshi_ceaser_2018} randomizes the location of lines in the last-level cache (LLC), and only claims to mitigate conflict-based cache attacks. DAWG \cite{kiriansky_dawg_2018} partitions caches into protection domains. On the more extreme side, Tsai \textit{et al.} \cite{tsai_rethinking_2018} redesign the memory hierarchy to replace caches with a memory-safe alternative they call Hotpads. While eliminating caches would eliminate cache-based side-channels, it does not protect against other side-channels, and the authors did not verify whether Hotpads might be used as side-channels.

One group of mitigations, which came to be known as \textit{invisible speculation}, focused on hiding changes to the cache. Yan \textit{et al.} \cite{yan_invisispec_2018}, Khasawneh \textit{et al.} \cite{khasawneh_safespec_2019}, Sakalis \cite{sakalis_ghost_2019},  Gonzalez \textit{et al.} \cite{gonzalez_replicating_2019}, Ainsworth and Jones \cite{ainsworth_muontrap_2020}, and Wu and Qian \cite{wu_rcp_2022} added a small separate cache to store speculative loads. Sakalis \textit{et al.} \cite{sakalis_efficient_2019} proposed delaying updates to the cache hierarchy until after a load is no longer speculative, so L1 data cache hits would execute speculatively, but L1 data cache misses would delay until they could execute non-speculatively.  Behnia \textit{et al.} \cite{behnia_speculative_2021} and Fustos \textit{et al.} \cite{fustos_spectrerewind_2020} later demonstrated that invisible speculation approaches are not effective mitigations, because the delayed load introduces timing changes that can be observed in subsequent instructions that depend on the load, so the secret can be inferred even though the cache hierarchy was not immediately updated. Even worse, the subsequent dependent instructions may update the cache, making the secret easily accessible through the cache anyway, despite the delayed load. GhostMinion \cite{ainsworth_ghostminion_2021} was proposed to resolve the security problems with previous approaches to invisible speculation, however Yang \textit{et al.} \cite{yang_pensieve_2023} uncovered a new variant of Spectre that bypasses GhostMinion.

CleanupSpec \cite{saileshwar_cleanupspec_2019}, ReversiSpec \cite{wu_reversispec_2020}, ReViCe \cite{kim_revice_2020}, and CacheRewinder \cite{lee_cacherewinder_2022} all take an approach of cleaning up the cache after speculation fails. However, all rollback techniques permit speculative execution to change the cache system, so they have the same problems as invisible speculation \cite{behnia_speculative_2021, fustos_spectrerewind_2020}, because the cache changes can still be observed by correct instructions executing at the same time as the misspeculated instructions, and the leakage succeeds before the cleanup finishes \cite{ainsworth_ghostminion_2021, wu_rcp_2022, guarnieri_hardware-software_2021}.

\subsubsection{Mitigation approaches based on isolation}\label{sec:spectre-mitigation-isolation}

Another general approach to mitigating Spectre has been to increase isolation between user and kernel modes, threads, processes, or other security domains. One problem with isolation approaches to mitigating Spectre is that flushing or partitioning some microarchitectural state when changing domains is generally not sufficient to eliminate all microarchitectural traces, and so hardware remains vulnerable despite the mitigations \cite{canella_evolution_2020}. A more fundamental problem with all mitigation approaches that rely on isolating one security domain from another is that not all Spectre variants are cross-domain attacks. Even the very first paper on Spectre \cite{kocher_spectre_2018} highlighted the fact that Spectre attack code could be in the same process and the same privilege level as the victim code, with a target of leaking memory that the attacker should not have access to because of a sandboxed interpreter, JIT compiler, or memory-safe language. Canella \textit{et al.} \cite{canella_systematic_2019} demonstrated that Spectre-PHT, Spectre-BHB, and Spectre-RSB variants still succeeded on Intel processors no matter whether the mistraining was done in the same-process or cross-process, and using the victim branch or a congruent branch. The same-process and cross-proces variants mostly succeeded on AMD and ARM too, though they both had some protections against cross-process congruent branch mistraining for Spectre-BTB, and ARM had some protection against cross-process mistraining for Spectre-RSB. Mitigations that merely isolate predictors across user/kernel mode or between threads are not effective against same-domain Spectre attacks \cite{koruyeh_spectre_2018, barberis_branch_2022}.

Intel and AMD added Indirect Branch Restricted Speculation (IBRS) \cite{noauthor_speculative_2018, noauthor_software_2023} as a hardware defense against Spectre-BTB, to flush branch predictor state when switching between user and kernel mode. Mambretti \textit{et al.} \cite{mambretti_two_2019} observed that IBRS was not effective against their icache and Double BTI variants of Spectre-BTB. Intel and AMD later added enhanced IBRS (eIBRS) as an improvement to IBRS, however Barberis \textit{et al.} \cite{barberis_branch_2022} demonstrated that eIBRS was not effective because it only protected the Branch Target Buffer (BTB), so the mitigation could be bypassed by mistraining the Branch History Buffer (BHB) instead. ARM added similar features in the form of ``IbrsSameMode'' and CSV2 features, which were also vulnerable to the BHB variant of Spectre \cite{barberis_branch_2022}. Furthermore, Barberis \textit{et al.} \cite{barberis_branch_2022} discovered variants of Spectre-BTB using same-mode indirect branch mispredictions (kernel-to-kernel), so that eIBRS and other isolation-based mitigations in general are not sufficient protection.

Intel and AMD added Single Thread Indirect Branch Predictors (STIBP) \cite{noauthor_speculative_2018, noauthor_software_2023} to isolate branch predictors for different hardware threads on the same core, preventing branch predictors in one thread from influencing branch predictions in other threads. STIBP has been reported to be effective as a partial mitigation only for cross-thread mistrainings \cite{mambretti_two_2019}, with performance penalties in the range of 50\% \cite{carna_fight_2022}, and both Intel and AMD have recommended against enabling STIBP by default \cite{corbet_taming_2018}. STIBP has no effect on co-resident processes \cite{wampler_exspectre:_2019} or other same-domain mistrainings.

Wistoff \textit{et al.} \cite{wistoff_microarchitectural_2021, wistoff_systematic_2022} proposed a \texttt{fence.t} instruction to provide temporal partitioning, and Escouteloup \textit{et al.} \cite{escouteloup_under_2022} proposed thread-level security domains called ``domes'', to introduce additional levels of isolation on RISC-V processors, but neither approach has any effect on same-domain attacks.

\subsubsection{Mitigation approaches based on selective speculation}\label{sec:spectre-mitigation-selective}

The most successful mitigation approaches to Spectre, in terms of both security and performance, have turned out to be the ones that restrict speculation. These approaches are based on a growing understanding that truly mitigating Spectre at the transmission phase (the side-channel leakage attack vector) would require blocking all known microarchitectural side-channels as well as any that might be discovered in the future \cite{green_safebet_2023}. Identifying the initial speculative access of the secret is a more tractable problem than chasing down every possible secondary transmission channel. And, once you have identified which instructions are risky to speculate, it is easier to prevent speculative execution than to chase down all the side effects after speculative execution has already happened.

As a mitigation for Spectre-STL, Intel introduced a processor mode Speculative Store Bypass Disable (SSBD) \cite{noauthor_speculative_2018}, which prevents loads from executing if they bypass any stores, so attackers cannot read stale values, effectively turning all loads non-speculative. While this mitigation reportedly works for Spectre-STL variants, it has no effect on other Spectre variants. Initially measured at an 8\% performance penalty in 2018 \cite{weisse_nda:_2019}, Behrens \textit{et al.} \cite{behrens_performance_2022} observed that grew to a 34\% performance penalty by 2022, possibly because newer processors may be shipping more complete implementations of SSBD than was possible with the original microcode patches. SSBD is defeated by other transient execution vulnerabilities such as RIDL \cite{van_schaik_ridl_2019}.

Some approaches to selective speculation simply delay the execution of all instructions that may have speculative sources of data as operands. NDA \cite{weisse_nda:_2019} restricts data propagation after an unresolved branch or unresolved store address. It begins with the assumption that instructions can execute speculatively as long as their operands are the results of ``safe'' instructions. They regard any instruction following a branch instruction as unsafe until the branch target and direction is resolved, and any load instruction as unsafe if it follows a store with an unresolved address. NDA then delays execution of any instruction with unsafe operands until those operands can be marked as safe, because the original speculation trigger for that operand has resolved and is no longer speculative. SpecShield \cite{barber_specshield_2019} is similar to NDA, but focuses more on load instructions as sources of speculative data forwarding, and makes some minimal attempt at identifying which instructions are a lower risk for leaking forwarded speculative data. NDA has performance penalties as high as 45\%, and SpecShield 21\%. Jin \textit{et al} \cite{jin_specterminator_2022} attributed the poor performance of both approaches to the way they delay execution of a large number of instructions that never could have caused changes to the microarchitectural state anyway, and so would have been safe to execute speculatively.

Some approaches to selective speculation are based on hardware taint tracking techniques, inspired by previous work on information flow tracking techniques \cite{tiwari_complete_2009, tiwari_execution_2009}. Speculative Taint Tracking (STT) \cite{yu_speculative_2019} begins with the assumption that it is safe to speculatively execute instructions and speculatively forward their results to other instructions, as long as: 1) the forwarded results are marked as ``tainted''; 2) the taint propagates as the forwarded results are used as operands for subsequent instructions; and 3) any tainted operands delay the execution of instructions that could serve as a transmission side-channel until the original instruction that tainted the operand is no longer speculative. STT was only proposed as a mitigation for Spectre-PHT variants, at a reported performance penalty of 14.5\%, however Loughlin \textit{et al.} \cite{loughlin_dolma_2021} later measured the performance penalty of STT as high as 44.5\% for protecting data in memory, and as high as 63.4\% when extended to protect data in registers. One key challenge of the STT approach is identifying all the instructions that could be used as transmission side-channels, and Jin \textit{et al} \cite{jin_specterminator_2022} and Loughlin \textit{et al.} \cite{loughlin_dolma_2021} later identified that STT does not catch Spectre variants using speculative store instructions in the transmission phase with side-channels based on the TLB, store buffer, or load-store aliasing. Choudhary \textit{et al.}  \cite{choudhary_speculative_2021} observed that STT only prevented speculative transmission of data that was accessed speculatively, but failed to protect data that was originally accessed non-speculatively, so their Speculative Privacy Tracking (SPT) extends the idea of STT, by tainting the results of a much larger set of data access instructions, with a performance penalty as high as 45\%. Speculative Data-Oblivious Execution (SDO) \cite{yu_speculative_2020} extended STT by allowing some transmission side-channel instructions to execute speculatively if they are independent of sensitive data, at a reported performance penalty of 10\%. Zhao \textit{et al.} \cite{zhao_pinned_2022} and Kvalsvik \textit{et al.} \cite{kvalsvik_doppelganger_2023} tried to improve the performance of STT and other similar approaches, by altering the behavior of speculative loads, reporting performance penalties of 13.2\% and 4.9\% respectively for their implementations of STT.

Dolma \cite{loughlin_dolma_2021} is conceptually similar to STT, but instead of taint tracking forwarded results of prior instructions, it tracks speculative control dependencies (on prior branch instructions) and speculative data dependencies (on prior load instructions). Dolma marks micro-ops in the reorder buffer with the speculative control or data dependency, and delays their execution until the dependency is resolved because the original store or load is no longer speculative. Dolma reported a performance penalty of 42.2\%, and claimed to protect against all transient execution attacks, but Jin \textit{et al} \cite{jin_specterminator_2022} noted that Dolma does not protect against a load-load reordering side channel, as identified by Yu \textit{et al.} \cite{yu_speculative_2019}. Conditional Speculation \cite{li_conditional_2019} also tracks dependencies on prior branch and load instructions like Dolma, however it only delays execution of potential transmission side-channel instructions if they would change cache contents due to mis-specuation because of a cache miss. This more limited approach lowers the performance penalty to 12.8\%, but still leaks information on cache hits \cite{guarnieri_hardware-software_2021} and with side-channels other than cache \cite{jin_specterminator_2022}.

Ravichandran \textit{et al.} \cite{ravichandran_pacman_2022} noted that mitigations based on information flow tracking such as STT, NDA, and Dolma only consider load instructions as the source of the taint, so they are not effective against variants of Spectre where the speculative taint has a different source, such as a pointer authentication instruction. The SpecHammer \cite{tobah_spechammer_2022} variant of Spectre-PHT defeats some taint tracking mitigations by using Rowhammer to flip bits in the victim code, so code that would not ordinarily work for the access phase of Spectre-PHT becomes a viable attack vector.

SpecTerminator \cite{jin_specterminator_2022} refines earlier selective speculation approaches with performance improvements to taint tracking, and by applying different delayed execution techniques to different kinds of sensitive instructions---TLB request ignoring, extended Delay-on-Miss, delayed squash, and selective issue. SpecTerminator considers side-channels based on the TLB, DRAM, BTB, and port contention in addition to cache-based side channels. Similar to other selective speculation approaches, SpecTerminator uses taint tracking for potential transmission side-channel instructions (loads or stores) that depend on earlier access instructions (loads). But, instead of only delaying execution of transmission instructions, they delay TLB requests, which blocks more potential side-channels at an earlier stage of the pipeline. This approach also delays issue of branch instructions that depend on a prior speculative load, to prevent speculative updates to other microarchitectural states that enable BTB or port contention side channels. And, this approach delays squashes to protect against Spectre-STL variants and load-load reordering. SpecTerminator reported an impressive 6\% performance penalty for mitigating the subset of Spectre variants they considered. However, Ghaniyoun \cite{ghaniyoun_short_2023} independently evaluated the SpecTerminator implementation and measured the performance penalty at 25\%---significantly higher than the 6\% reported in the original paper---and determined that TLB requests were not being ignored as intended.

SafeBet \cite{green_safebet_2023} focuses on the access phase of a Spectre attack, and delays execution of data access instructions until they are non-speculative. To improve performance, the approach uses a Speculative Memory Access Control Table (SMACT) to track prior non-speculative data accesses within the code region of a trust domain, and allows speculative data access instructions to execute if they are accessing the same location in the same region as a prior non-speculative data access. The SafeBet paper claims to mitigate all variants of Spectre, but then goes on to say the approach does not handle side channels based on micro-op caches. The approach only considers load instructions as sources of speculative data, so the limitation that Ravichandran \textit{et al.} \cite{ravichandran_pacman_2022} identified for STT, NDA, and Dolma would also apply to SafeBet. And, SafeBet is fundamentally an isolation mitigation, so it offers no protection against same-domain Spectre variants, as discussed in Section \ref{sec:spectre-mitigation-isolation}.

The greatest challenges for selective speculation mitigation approaches is determining where speculation is safe or unsafe, and how to disable speculation with the least possible disruption to legacy software stacks while providing strong security guarantees. Manual approaches are possible---leaving the decision of whether speculation is safe or unsafe to the software or compiler developer---but they can never provide strong security guarantees. The approaches described in this section are more automated---the pipeline makes all the decisions about where speculation is safe or unsafe. So far, these automated approaches still have not managed to provide strong security guarantees, because they miss some scenarios where speculation is unsafe or because the implementation fails to disable speculation where the design intended. But, over time selective speculation approaches have been getting closer to providing a comprehensive solution to Spectre with strong security guarantees. There may be room for a middle-ground selective speculation approach that provides strong security guarantees by disabling speculation for a security domain---such as a container, VM, secure enclave, serverless function, or small region of code---to protect code within the security domain from both cross-domain transient execution attacks launched outside the security domain and same-domain attacks launched within the security domain, and also serve as a sandbox preventing code inside the security domain from launching cross-domain attacks on any other part of the system.

%% file: 04_meltdown.tex
\section{Meltdown}

Like Spectre, Meltdown is a transient execution vulnerability first discovered in 2017 and reported publicly in January 2018, by Lipp \textit{et al.} \cite{lipp_meltdown_2018}, in a preprint which was republished later that year at the USENIX Security Symposium in June \cite{lipp_meltdown_2018-1}. Also like Spectre, Meldtown is a fault analysis side-channel attack---it combines both fault-injection techniques to manipulate the victim into a vulnerable state and side-channel techniques to convey the exposed secrets to the attacker. Unlike Spectre, Meltdown does not use speculation as an attack vector, so an out-of-order pipeline can be vulnerable to Meltdown, even if it has no speculative features.

Research on Meltdown variants and mitigations has been far less extensive than Spectre, probably partly due to the fact that AMD, ARM, and IBM processors were never vulnerable to some variants of Meltdown \cite{canella_systematic_2019, genkin_whack--meltdown_2021}, so we have always known that hardware mitigations for Meltdown could have reasonable security and performance. Eventually, even Intel figured out that faulty reads could just return zero, preventing the leak of secret information \cite{genkin_whack--meltdown_2021}.

\subsection{Characterizing the variants}

A number of variants of Meltdown have been reported, primarily focused on unauthorized access to some value protected by a permission check, and the defining characteristics of all variants are two phases: 1) triggering an exception for a failed permission check in the context of transient execution so the exception is delayed; and 2) leaking the unauthorized value through microarchitectural side channels. The permission check will ultimately fail and raise an exception, but in the context of transient execution, the exception is delayed until the transient instruction sequence commits. Some microarchitecture implementations have historically made the design choice to update shared microarchitectural state during transient execution as if the permission checks were successful, and to allow subsequent transient instructions in the sequence to operate using the unauthorized value. In theory, those changes are only temporary and never architecturally visible, but in practice, shared microarchitectural state can be observed by an attacker and leaked over side channels.

The primary way of categorizing Meltdown-type variants, shown in Table \ref{table:meltdown-variants-by-exception}, is by the exception used in the preparation phase. A secondary way of categorizing Meltdown-type variants is by the microarchitectural states used in the transmission phase of attack---Table \ref{table:meltdown-variants-by-channel} shows some of the highlights. There have been fewer attempts to replicate Meltdown variants across a diverse collection of different side channels, because it quickly became clear that it was feasible to block Meltdown in the preparation and access phases of the attack, so the side channel used in the transmission phase is less interesting.

\begin{table*}[h!]
	\small
	\caption{Meltdown variants by exception}
	\label{table:meltdown-variants-by-exception}
	\begin{tabular}{p{2.5cm} p{3cm} p{5cm} p{4cm}}
		\toprule
		\textbf{Exception} & \textbf{Permission Bit} & \textbf{Mechanisms} & \textbf{Examples} \\
		\midrule
		
		page fault & user/supervisor page-table attribute & Supervisor-only Bypass: bypasses user/supervisor permission checks to read unauthorized kernel memory from user space. & Meltdown (original variant, ``Rogue Data Cache Load'') \cite{lipp_meltdown_2018, lipp_meltdown_2018-1}\\
		\midrule
		
		page fault & read/write page-table attribute & Read-only Bypass: bypasses read/write permission checks to transiently write over read-only data within the current privilege level. May be used, for example, to bypass the hardware-enforced isolation of software-based sandboxes. & Meltdown-RW (also inaccurately called ``Spectre variant 1.2'') \cite{kiriansky_speculative_2018, canella_systematic_2019}\footnote{Note that although Meltdown-RW is generally categorized as a Meltdown-type attack, it is not a side-channel attack because the target of the attack is not to leak secret values (violating confidentiality), but instead to modify read-only values (violating integrity).}\\
		\midrule
		
		page fault & page-table present bit or reserved bit & L1 Terminal Fault (L1TF): bypasses Intel SGX enclave or operating system or hypervisor isolation to read unauthorized memory across isolation boundaries. & Foreshadow (Intel SGX) \cite{van_bulck_foreshadow:_2018}, Foreshadow-NG (OS and hypervisor) \cite{weisse_foreshadow-ng:_2018}, Foreshadow-VMM (VM guest to host) \cite{brunella_foreshadow-vmm_2019}\\
		\midrule
		
		page fault & Intel memory-protection keys for user space (PKU) & Protection Key Bypass: bypasses hardware-enforced read and write isolation, to leak or modify protected memory. & Meltdown-PK \cite{canella_systematic_2019}\\
		\midrule
		
		page fault & not present, all access to the page has been revoked & Write Transient Forwarding (WTF): store buffer & Fallout \cite{minkin_fallout_2019}\\
		\midrule
		
		general protection fault & \textit{N/A} & System Register Bypass: bypasses permission checks on privileged system registers to leak system register contents. & Meltdown-GP (also called variant 3a) \cite{canella_systematic_2019}\\
		\midrule
		
		device not available exception & \textit{N/A} & FPU Register Bypass: bypasses isolation of floating point unit or SIMD registers across context switches, to leak register contents. & Lazy FP \cite{stecklina_lazyfp_2018}\\
		\midrule
		
		bound range exceeded exception & \textit{N/A} & Bounds Check Bypass: bypass hardware-enforced array bounds checking\footnote{Such as the older Intel \texttt{bound} opcode or modern Intel Memory Protection eXtensions (MPX).} to access out-of-bound array indices. & Meltdown-BR \cite{dong_spectres_2018, canella_systematic_2019} including Meltdown-MPX \cite{noauthor_intel_2018} and Meltdown-BND \cite{canella_systematic_2019}\\
		\bottomrule

	\end{tabular}
\end{table*}

\begin{table*}[h!]
	\small
	\caption{Meltdown variants by transmission phase side-channel attack vector}
	\label{table:meltdown-variants-by-channel}
	\begin{tabular}{p{4cm} p{5cm} p{5.9cm}}	
		\toprule
		\textbf{Channel} & \textbf{Mechanisms} & \textbf{Examples} \\
		\midrule
		
		L1 data cache & Leaks information using a cache-timing side channel on the L1D cache & L1TF variants \cite{van_bulck_foreshadow:_2018, weisse_foreshadow-ng:_2018, brunella_foreshadow-vmm_2019} and SMAP and MPK variants \cite{xiao_speechminer_2020} only work on L1D\\
		\midrule
		
		L3 cache or LLC & For example, Flush+Reload \cite{yarom_flush+reload:_2014} or Prime+Probe \cite{liu_last-level_2015} & Meltdown (original variant) \cite{lipp_meltdown_2018, lipp_meltdown_2018-1}, Meltdown-GP (also called variant 3a) \cite{canella_systematic_2019}, Meltdown-PK \cite{canella_systematic_2019}, Lazy FP \cite{stecklina_lazyfp_2018}\\
		\midrule
		
		uncached memory & Leaks information using a DRAM-based side channel & Meltdown (original variant) \cite{lipp_meltdown_2018, lipp_meltdown_2018-1}\\
		\midrule
		
		Translation Lookaside Buffer (TLB) & Leaks information using a TLB-based side channel & Schwarz \textit{et al.} \cite{schwarz_store--leak_2021}, Seddigh \textit{et al.} \cite{seddigh_breaking_2022}\\
		\bottomrule

	\end{tabular}
\end{table*}

While AMD was not vulnerable to earlier variants of Meltdown, it was vulnerable to the Meltdown-BND variant \cite{canella_systematic_2019} in Table \ref{table:meltdown-variants-by-exception} and to new variants reported by Xiao \textit{et al.} \cite{xiao_speechminer_2020} in Table \ref{table:meltdown-variants-by-channel}.

\subsection{Characterizing the countermeasures}\label{sec:meltdown-countermeasures}

A number of different countermeasures were proposed for Meltdown-type attacks, but ultimately the right answer was fairly simple: always do permission checks first, and never update shared microarchitectural state or forward the results of data accesses until after the permission checks are successful \cite{yan_invisispec_2018, genkin_whack--meltdown_2021}. It is fine to delay raising the exception until after the transient instructions commit, so out-of-order and speculative pipelines can be safe from Meltdown-type vulnerabilities as long as the microarchitecture design is done correctly. The only reason Meltdown-type attacks ever worked, is that hardware designers assumed that microarchitectural state created in the context of transient execution was safely hidden so deep in the hardware that it could never be accessed, but that assumption was false.

There were some early software-only mitigations for Meltdown, which are still in use on legacy hardware. The KAISER \cite{gruss_kaslr_2017} patch to Kernel Address Space Randomization (KASLR) was demonstrated to be an effective mitigation for the first User/Supervisor variant of Meltdown \cite{lipp_meltdown_2018}, and was later implemented in the Linux Kernel as Kernel Page Table Isolation (KPTI) \cite{corbet_page-table_2018}. However, KAISER and KPTI are only isolation mitigations between kernel and user space memory, and so the mitigation has no effect on other variants of Meltdown or on same-mode attacks. Hua \textit{et al.} \cite{hua_epti_2018} measured the KPTI mitigation at a 30\% performance penalty, and developed an alternative mitigation, EPTI, that uses extended page tables (EPT) instead of guest page tables for isolation at a 13\% performance penalty. While EPTI performed better than KPTI, it was not more effective. Page Table Entry (PTE)-Inversion \cite{corbet_meltdown_2018} was implemented as a mitigation for the L1 Terminal Fault (L1TF) variants of Meltdown, by ensuring that addresses used following a translation failure do not point to a valid page frame \cite{genkin_whack--meltdown_2021}. He \textit{et al.} \cite{he_new_2021} observed that software-only mitigations have been far less successful for Meltdown than they were for Spectre, because the microarchitectural causes for Meltdown-type vulnerabilities occur within a single instruction, while the microarchitectural causes for Spectre-type vulnerabilities occur in the interaction between instructions.

Isolation mitigations were also tried, such as flushing the L1 cache on context switches or careful scheduling to prevent processes or VMs from executing on the same core or thread \cite{weisse_foreshadow-ng:_2018, genkin_whack--meltdown_2021}. And, a number of mitigations for Spectre also claimed to mitigate Meltdown, with varying degrees of success \cite{weisse_nda:_2019, khasawneh_safespec_2019, green_safebet_2023}, even though Meltdown-type attacks really are fundamentally different than Spectre-type attacks \cite{he_new_2021}. The proliferation of hardware and software mitigations necessary to catch all variants of Meltdown have been deeply unappealing compared to AMD's simple answer of ``just don't be vulnerable in the first place'' \cite{weisse_nda:_2019, genkin_whack--meltdown_2021, nosek_evaluation_2022}.

However, just because it is possible to eliminate Meltdown-type vulnerabilities from out-of-order and speculative cores with careful microarchitecture design, does not mean that every microarchitecture implementation has successfully done so. This is one of many reasons why pre- and post-silicon hardware security verification techniques are critical for modern hardware design, as discussed in Section \ref{sec:verification}.

%% file: 05_beyond.tex
\section{Transient execution vulnerabilities beyond Spectre and Meltdown}\label{sec:beyond-spectre-meltdown}

Because Spectre and Meltdown were the first transient execution vulnerabilities discovered, they have received the most attention, but researchers continue to find new transient execution vulnerabilities. The vulnerabilities all share the defining characteristic of using transient execution effects as an attack vector, but otherwise they are a diverse collection. Some are side-channel attacks with a goal of leaking secrets to violate confidentiality like Spectre and Meltdown, but others are straight up fault-injection attacks with a goal of violating integrity.

\subsection{Side-channel attacks inspired by Meltdown}

Some transient execution vulnerabilities use different ways of inducing transient execution. Rather than exploiting delayed exceptions like Meltdown, Nemesis \cite{van_bulck_nemesis:_2018} exploits the fact that interrupts are delayed until instruction retirement. The target of Nemesis-type attacks is to leak instruction timings from secure enclaves. Fallout \cite{minkin_fallout_2019} uses microcode assists as a trigger for transient execution rather than exceptions, leaks information via the store buffer, and is able to bypass the Kernel Page Table Isolation (KPTI) countermeasure for Meltdown. 

Possibly inspired by an early mention of line-fill buffers as a potential attack vector for Meltdown \cite{lipp_meltdown_2018-1}, microarchitectural data sampling (MDS) attacks are not triggered by either exceptions (like Meltdown) or speculative predictions (like Spectre), but instead exploit the transient effects of line-fill buffers, load ports, and store buffers. Rogue In-flight Data Load (RIDL)  \cite{van_schaik_ridl_2019} cannot be mitigated in software, and specifically defeats mitigations such as Kernel Page Table Isolation (KPTI), Page Table Entry (PTE) inversion, Speculative Store Bypass Disable (SSBD), and L1 data cache flushing, and works both cross-context and same-context. ZombieLoad  \cite{schwarz_zombieload_2019} amplifies microarchitectural data sampling (MDS) and bypasses mitigations for both Meltdown-type attacks and other MDS-type attacks. CacheOut \cite{van_schaik_cacheout_2020} bypasses mitigations that Intel put in place on the Whiskey Lake architecture to protect against other MDS-type attacks such as Fallout, ZombieLoad, and RIDL. SGAxe \cite{van_schaik_sgaxe_2020} adapts CacheOut to target SGX enclaves. Medusa \cite{moghimi_medusa_2020} is a more focused MDS-type attack than ZombieLoad or RIDL, which only targets data loads caused by write combining operations, and can only be successfully mitigated if hyperthreading is disabled. Ragab \textit{et al} \cite{ragab_crosstalk_2021} discovered another variant of an MDS-type attack that leaks information using a global staging buffer shared between all CPU cores and defeats mitigations based on spatial or temporal partitioning or isolating workloads on separate cores. Witharana and Mishra \cite{witharana_speculative_2022} reported another MDS variant that works on AMD architectures, which were not vulnerable to previous variants.

The Gather Data Sampling (GDS) \cite{moghimi_downfall_2023} attack exploits the x86 \texttt{gather} instruction in the context of transient execution to leak stale data from the shared SIMD register buffers.

\subsection{Side-channel attacks inspired by Spectre}

Rokicki \cite{rokicki_ghostbusters_2020} demonstrated that processors based on Dynamic Binary Translation (DBT), such as Nvidia Denver \cite{boggs_denver_2015} or Hybrid-DBT \cite{rokicki_hybrid-dbt_2019}, are vulnerable to variants of Spectre even though the underlying hardware is strictly in-order, because the DBT engine introduces conditional branch prediction and memory dependency prediction as it translates and optimizes the binaries.

\subsection{Other transient execution vulnerabilities}

Not all transient execution vulnerabilities are side-channel attacks, some use transient execution effects for other purposes. Like Meltdown, Load Value Injection (LVI) \cite{van_bulck_lvi_2020, easdon_rapid_2022} begins with a preparation phase of triggering an exception, but the target of the attack is fault-injection rather than side-channel leakage, specifically to inject false values into the victim's transient execution (violating integrity). Also, LVI attacks run in the victim domain, so cross-domain isolation is not effective as a mitigation \cite{canella_evolution_2020}. The Gather Value Injection (GVI) \cite{moghimi_downfall_2023} attack extends LVI using the Gather Data Sampling (GDS) technique, with the same target of value injection.

Ragab \textit{et al.} \cite{ragab_rage_2021} explored transient execution vulnerabilities on Intel and AMD induced by machine clears, rather than mispredictions like Spectre or delayed exceptions like Meltdown. Their Speculative Code Store Bypass (SCSB) variant allows attackers to execute stale code, while their Floating Point Value Injection (FPVI) variant is similar to LVI but injects operands into floating point operations. Both are primarily integrity attacks, but they can also be combined with side-channel attack techniques (violating confidentiality).

Like Spectre, ExSpectre \cite{wampler_exspectre:_2019} has a preparation phase that mistrains branch predictors, but unlike Spectre, it uses transient execution effects to hide malware from static and dynamic analysis techniques, with a primary target of arbitrary code execution (violating integrity). For example, ExSpectre is  capable of running system calls to launch a dial-back TCP shell. Isolation techniques such as Intel's Single Thread Indirect Branch Predictors (STIBP) are not effective mitigations against ExSpectre because the attack code and the victim code run in the same context.

GhostKnight \cite{zhang_ghostknight_2022} has a preparation phase that mistrains branch predictors, but uses speculation execution to amplify the Rowhammer fault-injection attack, extending the reach of the attack to cross privilege boundaries (violating integrity). Spoiler \cite{islam_spoiler_2019} also uses transient execution effects to amplify Rowhammer attacks.

BlindSide \cite{goktas_speculative_2020} is a speculative probing technique that uses speculative execution to amplify a simple memory corruption attack into a speculative control-flow hijacking attack, with targets ranging from leaking sensitive data, to arbitrary code execution, all the way to full-system compromise. Speculative probing attacks are able to bypass mitigations designed to prevent speculative control-flow hijacking such as retpoline, IBPB, IBRS, and STIBP.

Another category of vulnerabilities that can use transient execution effects are microarchitectural replay attacks (MRA) such as MicroScope \cite{skarlatos_microscope_2019, skarlatos_jamais_2021, sakalis_delay--squash_2022}, where the attacker forces pipeline flushes so the victim instructions are repeatedly re-executed. MRA techniques can reduce the noise in side channels used to leak secrets, making transient execution vulnerabilities and other vulnerabilities easier to exploit.

%% file: 06_verification.tex
\section{Hardware security verification for transient execution}\label{sec:verification}

Over the years of research into the transient execution vulnerabilities, the emphasis has shifted away from looking for some magic hardware or software countermeasure that will preserve the performance benefits of transient execution while eliminating the security risks. Instead, there is a growing understanding of transient execution as one of those complex multilayered problems, like memory safety, where human errors by the people designing and implementing the systems plays a significant role, and expecting hardware engineers to manually catch all the security flaws is an inadequate answer. In response---and as part of a broader trend of increasing interest in hardware security verification \cite{erata_survey_2023, anton_fault_2023, pearce_high-level_2023, canakci_processorfuzz_2023, witharana_survey_2022, kulik_survey_2022, hu_overview_2021, hu_hardware_2021, dessouky_hardfails_2019, ferraiuolo_verification_2017, hu_towards_2016}---there has been a rise in academic and commercial tools to inspect, test, fuzz, and scan for transient execution vulnerabilities, at the hardware-level, at the software-level, or with formal models.

Hardware security verification tools are not capable of guaranteeing that a speculative or out-of-order processor is invulnerable to all transient execution vulnerabilties, but they can help improve security by determining whether a specific processor is vulnerable to specific known variants, confirming whether implemented and deployed mitigations actually work, and identifying risky patterns in the design and implementation of hardware. If you are a hardware vendor, the formal and pre-silicon tools can help you detect and fix flaws in your microarchitecture design and implementation before an expensive tape-out, the post-silicon tools can help ensure that the security features you designed work as intended in physical form, and the software-only tools can be helpful in hardware enablement efforts to ensure that software your customers are likely to run works well on your hardware and benefits from your security features. If you are a software developer, the post-silicon and software-only tools can help you discover how secure your hardware really is, and what adaptations you might need to make to protect your software and your users. 

Among the major hardware vendors, we know from publicly available information that ARM has used hardware security verification tools for the transient execution vulnerabilities \cite{madhu_enabling_2022}. Intel and AMD have been less forthcoming about the tools they use internally, however based on available evidence---specifically the way that AMD was not vulnerable to several variants of Meltdown and Spectre before they were even reported---it seems likely that AMD uses microarchitecture-level hardware security verification tools.

\subsection{Formal model verification}

Spectector \cite{guarnieri_spectector_2019} was an early attempt at detecting Spectre vulnerabilities using symbolic execution and comparing the microarchitectural information flows between speculative and non-speculative execution. Loughlin \textit{et al.} \cite{loughlin_dolma_2021} argued that Spectector was too restrictive and delayed some transient instructions that would have been safe to execute speculatively. Guarnieri \textit{et al.} \cite{guarnieri_hardware-software_2021} extended Spectector with a concept of speculation contracts. Fabian \textit{et al.} \cite{fabian_automatic_2022} extended Spectector beyond modeling branch instructions to also model store and return instructions, so it could detect variants of Spectre-PHT, Spectre-RSB, and Spectre-STL.  CacheFix \cite{chattopadhyay_symbolic_2018} and CheckMate \cite{trippel_checkmate_2018} both do formal modeling of microarchitectural state to detect vulnerabilities, but only for cache-timing side channel attacks.

Cauligi \textit{et al.} \cite{cauligi_sok_2022} surveyed formal frameworks for software mitigations for Spectre. Cheang \textit{et al.} \cite{cheang_formal_2019} formally defined a class of information flow security properties for reasoning about the security of microarchitectural speculation features, and operational semantics for an intermediate assembly representation which can run small programs and verify if they conform to the secure speculation property. Griffin and Dongol \cite{griffin_verifying_2021} implemented the secure speculation properties defined by Cheang \textit{et al.} in the Isabel/HOL proof assistant. Unique Program Execution Checking (UPEC) \cite{fadiheh_processor_2019,fadiheh_formal_2020,fadiheh_exhaustive_2023} applied a structured and systematic formal methodology for hardware security verification that targets transient execution vulnerabilities at the register-transfer level (RTL) of the hardware design and implementation workflow. InSpectre \cite{guanciale_inspectre_2020} proposed a formal microarchitectural model of out-of-order and speculative features used as attack vectors in a variety of transient execution vulnerabilities, and implements the model as  an abstract microcode target language for translating ISA instructions, Machine Independent Language (MIL). Pitchfork \cite{cauligi_constant-time_2020} performed constant-time code analysis on an abstract model, but lacks microarchitectural implementation details. KLEESpectre \cite{wang_kleespectre_2020} extended the KLEE symbolic execution engine with modeling of cache and speculative execution. 

Pensieve \cite{yang_pensieve_2023} formally modeled early-stage microarchitectural designs, to evaluate the security of proposed mitigations for transient execution vulnerabilities. Ponce-de-le\'on and Kinder \cite{ponce-de-leon_cats_2022} used the CAT modeling language for memory consistency to implement an axiomatic framework to detect attacks and validate defenses for transient execution vulnerabilities, including execution models of speculative control flow, store-to-load forwarding, predictive store forwarding, and machine clears.
Mathure \textit{et al.} \cite{mathure_refinement-based_2022} applied refinement-based formal verification methods to detect whether a microarchitecture design is vulnerable to variants of Spectre.

\subsection{Pre-silicon verification}

Hu \textit{et al.} \cite{hu_hardware_2021} surveyed hardware verification strategies based on information flow tracking, for a variety of hardware security vulnerabilities including the transient execution vulnerabilities.

Barber \textit{et al.} \cite{barber_pre-silicon_2022} instrumented RTL simulations to produce detailed execution traces of microarchitectural structures, and perform differential analysis on the traces to identify potential attack vectors. TEESec \cite{ghaniyoun_teesec_2023} is a pre-silicon framework for discovering microarchitectural vulnerabilities in secure enclaves, by profiling the processor design for microarchitectural structures relevant to enclave data, crafting verification gadgets to exercise all possible access paths to the enclave data, running the verification gadgets through a cycle-accurate RTL simulation of the design-under-test, and analyzing the simulation logs for traces that violate microarchitectural security principles.

SpecDoctor \cite{hur_specdoctor_2022} is an automated RTL fuzzer to detect both Spectre-type and Meltdown-type vulnerabilities, which systematically tests a comprehensive set of configuration options while selectively monitoring specific RTL components to discover constraint violations, then chains those violations to construct concrete proof-of-concept transient execution attack instruction sequences. SpecDoctor was implemented by adding monitoring logic for reorder-buffer rollback events to the Chisel source code for two RISC-V core implementations, BOOM and NutShell. IntroSpectre \cite{ghaniyoun_introspectre_2021} is another RTL fuzzer to detect Meltdown-type leaks.

\subsection{Post-silicon verification}

SpeechMiner \cite{xiao_speechminer_2020} is a software framework focused on detecting transient execution vulnerabilities on existing hardware, by generating sequences of instructions as tests. It models Meltdown-type vulnerabilities as a race condition between data fetching and processor fault handling and models Spectre-type vulnerabilities as a race condition between side-channel transmission and speculative instruction squashing. SpeechMiner was only implemented for 32-bit and 64-bit x86 architectures, not ARM or RISC-V. Revizor \cite{oleksenko_revizor_2022, oleksenko_hide_2023} is a black-box testing framework that detects microarchitectural leakage on x86 CPUs, using a concept of speculation contracts.

Transynther \cite{moghimi_medusa_2020} used fuzzing techniques to systematically identify whether hardware is vulnerable to variants of Meltdown and microarchitectural data sampling (MDS) attacks. Transynther was only implemented for x86 (Intel and AMD) and has not been ported to ARM or RISC-V. Osiris \cite{weber_osiris_2021} and SIGFuzz \cite{rajapaksha_sigfuzz_2023} are also fuzzing frameworks to detect microarchitectural side channels. Plumber \cite{ibrahim_microarchitectural_2022} is a framework that generates instruction sequences from templates to identify side-channel behavior, using concepts from instruction fuzzing, operand mutation, and statistical analysis. It was only implemented for ARM and RISC-V, but could be ported to x86. Scam-V \cite{buiras_validation_2021} generates tests to validate side-channel models, based on validation of  information flow properties using relational analysis.

Li and Gaudiot \cite{li_online_2018,li_detecting_2019}, Depoix and Altmeyer \cite{depoix_detecting_2018}, Ahmad \cite{ahmad_real_2020}, and Alam \textit{et al.} \cite{alam_victims_2021} used a combination of hardware performance counters and machine-learning classifiers to detect Spectre and Meltdown attacks, and more broadly cache side-channel attacks, in live running hardware. CloudShield \cite{he_cloudshield_2023} used similar techniques to detect Spectre, Meltdown, and cache-based side-channel attacks on server hardware deployed in a cloud infrastructure. However, Dhavlle \textit{et al.} \cite{dhavlle_cr-spectre_2022} demonstrated that these detection mechanisms can be bypassed by variants of Spectre that use same-domain code-injection as part of the attack, and Pashrashid \textit{et al.} \cite{pashrashid_fast_2022} demonstrated they can be bypassed by Spectre variants that chain benign gadgets or insert \texttt{nop} instructions into the branch mistraining code.

Spectify  \cite{pashrashid_fast_2022} tracks the attack phases of a Spectre attack using microarchitecture-level information to find and report data leaks before the transmission phase of the attack, to help identify where hardware mitgations for Spectre need to be applied.

ABSynthe \cite{gras_absynthe_2020} takes an automated, black box approach to synthesizing contention-based side-channel attacks for x86 and ARM microarchitectures, which can be used by hardware designers for regression testing. 

\subsection{Software-only mitigation verification}

Kasper \cite{johannesmeyer_kasper_2022} is a software scanner for the Linux Kernel that looks for code sequences that could be used as gadgets in the access phase of a Spectre-PHT attack, and models not only cache-based side channels, but also port-contention side channels, MDS-based side channels, and LVI. Kasper operates as a fuzzer on the syscall interface, and requires recompiling the kernel with support for the scanner. SpecFuzz \cite{oleksenko_specfuzz_2020} enhanced conventional fuzzing techniques with instrumentation to simulate speculative execution. FastSpec \cite{tol_fastspec_2021} used fuzzing and deep learning techniques to automatically generate and detect Spectre gadgets.

Mosier \textit{et al.} \cite{mosier_axiomatic_2022} developed a static analysis tool for software based on their concept of a microarchitectural leakage containment model (LCM), which is able to identify some Spectre vulnerabilities. RelSE \cite{daniel_hunting_2021} performs static analysis of program binaries for Spectre-PHT and Spectre-STL, based on security property of speculative constant-time.

The CrossTalk \cite{ragab_crosstalk_2021} framework analyses the microarchitectural behavior of x86 instructions, with special attention to their use of globally shared staging buffers. Easdon \textit{et al.} \cite{easdon_rapid_2022} developed two open source frameworks---Transient Execution Attack library (libtea) and SCFirefox---to generate prototype Meltdown, LVI, and MDS attacks on x86 and ARM.

%% file: 07_conclusion.tex
\section{Conclusion}

So far, the transient execution vulnerabilities have not been handled particularly well. That does not mean they are impossible to defeat, or that we should settle for the current untenable compromise of plugging a few leaks while leaving massive gaping holes of known vulnerabilities. What it does mean, is that we need to move beyond looking for a quick fix, and take the time to understand the true nature of the transient execution vulnerabilities, and the reasons why some countermeasures have been effective and others have not.

We cannot predict what new transient execution vulnerabilities and variants future research might discover, but we can observe patterns in the vulnerabilities we already know about, and extrapolate. One key pattern takes advantage of the permissive nature of transient execution---allowing instructions to execute and microarchitectural state to be created or modified in ways that would never happen in normal non-transient (non-speculative and in-order) execution---which enables powerful attack vectors for constructing a wide variety of vulnerabilities, including the ability to redirect control flow to chosen code, inject values and code, and access any shared microarchitectural state. Another key pattern takes advantage of unrestricted global sharing of predictions and other microarchitectural state in modern microarchitectures. Considering that unrestricted global sharing is a well-known security risk pattern across all levels of the hardware and software system stack, it is surprising that we ever thought we could get away with it at the microarchitecture level with no negative consequences. These patterns are the building blocks for future transient execution vulnerabilities, but they are also clues that can lead us to more effective countermeasures and more resilient microarchitecture designs.